\documentstyle[aas2pp4]{article} 
\def\pd#1#2{\displaystyle\frac{\partial#1}{\partial#2}}
\begin{document}

\title{A Model of the Double Magnetic Cycle of the Sun}
\author{E. E. Benevolenskaya}
\affil{Pulkovo Astronomical Observatory, St. Petersburg, 196140,
       Russia}
\affil{W.W.Hansen Experimental Physics Laboratory,\\
       Stanford University, Stanford, CA 94305-4085}

\begin{abstract}

It has been argued that the solar magnetic cycle consists of two
main periodic components: a low-frequency component (Hale's
22-year cycle) and a high-frequency component (quasi-biennial
cycle). The existence of the double magnetic cycle on the Sun is
confirmed using Stanford, Mount Wilson and Kitt Peak magnetograph
data from 1976 to 1996 (solar cycles 21 and 22). In the frame of
the Parker's dynamo theory a model of the double magnetic cycle is
presented. This model is based on the idea of two dynamo sources
separated in space. The first source of the dynamo action is
located near the bottom of the convection zone, and the second
operates near the top. The model is formulated in terms of two
coupled systems of non-linear differential equations. It is
demonstrated that in the case of weak interaction between the two
dynamo sources the basic features of the double magnetic cycle
such as existence of two component and observed temporal
variations of high-frequency component can be reproduced.
\end{abstract}


\section{\bf Introduction}

The 22-yr cycle of solar activity is a magnetic cycle which
consists of two 11-years sunspot cycles. The 22-yr cycles begin on
even sunspot cycles according to the Z\"urich numbering
(\cite{Gne48}) and manifest themselves in reversal of polarity of
sunspots (Hale's law). From one 11-yr cycle to another, the
polarity of the preceding ($p$) and following ($f$) sunspots
reverses. This reversal corresponds to the changing of the
toroidal magnetic field ($B_\varphi$-component). During
even-numbered cycles $B_\varphi > 0$ (the direction of the
magnetic vector coincides with the direction of solar rotation) in
the northern ($N$) hemisphere and $B_\varphi < 0$ in the southern
($S$) hemisphere. This relation is reversed during odd-numbered
cycles. In parallel with this, the poloidal magnetic field, or
back\-ground field ($B_r$-component), shows a 22-yr periodicity
(\cite{How81}), and other studies (\cite{Ste94}) have suggested
that the low-latitude and polar fields similarly show a 22-yr
periodicity . Furthermore, these studies have hinted at a shorter
term periodicity of about 2~yr (high-frequency component). In
addition, during the periods of the three-fold polar magnetic
field reversals which occur during some sunspot cycle maxima, the
temporal separation of the zones of alternated polarity of the
magnetic field on $H_\alpha$ charts is approximately equal to
1.5--2.5 years  (\cite{Wald73}, \cite{Ben91}). Similar
periodicities were found in variations of radio flux on $10.7 {\rm
cm}$ (\cite{Belm66}), flares and sunspot areas (\cite{Akio87}).
The high-frequency, 2-year, component is substantially weaker than
the main 22-year cycle and its intensity varies with time. The two
year component more clearly appeared in northern hemisphere in
cycle 20 and in southern hemisphere in cycle 21; its value was
smaller in cycle 22 (\cite{Ben96}). However, the biennial cycle
represents a challenge to the solar dynamo theories which usually
explain only the main cycle. In this Letter, I present a new
evidence of the two components of the solar cycle, and an attempt
to explain the results in term of  Parker's dynamo theory (\cite
{Par79}).

\section{\bf Analysis of magnetograph data}

Using Kitt Peak magnetograph data we show how synoptic (longitude
independent) magnetic field patterns evolve through cycles 21 and
22. Each of the synoptic maps is represented by the values of
$B_\Vert$ as a function sine latitude and Carrington longitude.
The Carrington coordinate system is a reference frame rotating
rigidly at the rate which corresponds to the synodic rotation rate
of sunspots at a latitude of about $16^o$. The observed
line-of-sight components ($B_\Vert$) averaging over all longitudes
for each Carrington Rotation  is represented in Figure~1b. The
time step in the time set of values $B_\Vert$ corresponds to
Carrington rotation period ($P= 27.2753$ days).

\begin{figure}
\epsscale{1}
\plotone{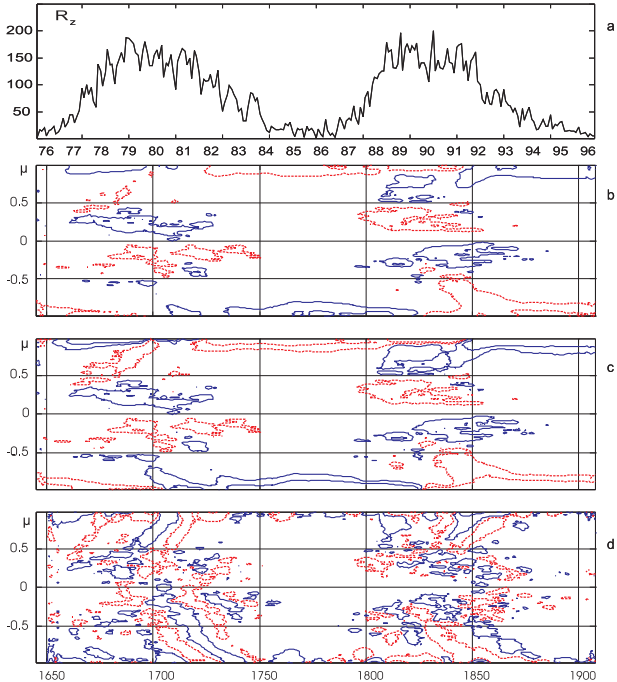}
\caption{
(a)~The values of the
sunspot number for cycles 21 and 22
             as function of time (in years).
(b)~The line-of-sight component of the solar magnetic field as function
     of Carrington rotation number.
(c)~The radial component of the solar magnetic field
    (blue lines correspond to isolines for +3, +10 $G$, red lines,
    for -3, -10 $G$).
(d)~The values of the $\Delta B_r /\Delta t$ for $\Delta t=6$
    solar Carrington rotations
    (blue lines correspond to isolines for +0.3, +3 $G$/rotation,
    red lines, for -0.3, -3 $G$/rotation).
 }
\end{figure}

The relative sunspot number is shown in Figure~1a.
Component $B_r$  (Figure~1c), was calculated from observational data
assuming that the true average field direction is radial and $B_\theta
 \cong 0$. The radial component will be easily found at  all latitudes
besides near the poles:

\begin{equation}
B_r=B_\Vert / \sin \theta,
\end{equation}

\noindent
where $\theta$ is colatitude.

To separate the high-frequency component in the data, we apply a
difference filter by computing $\Delta B_r /\Delta t$ for
different time intervals $\Delta t$. This is a reasonable
procedure if the poloidal magnetic field consists of two
components and  is represented as two dynamo waves:

\begin{equation}
B_r(t,\theta)=b_r(\theta)\sin\omega_ct +
              {b_r(\theta)\over A}\sin(\omega_\delta t+\varphi).
\label{field}
\end{equation}

\noindent where $b_r(\theta)$ is the amplitude of the
low-frequency radial component of the magnetic field; $A$ is ratio
between amplitudes of low-frequency and high-frequency components;
$\omega_c $ is the frequency of the Hale's solar cycle;
$\omega_\delta $ is the frequency of biennial cycle; $\varphi$ is
a phase.

The expression for the $\Delta B_r/\Delta t$ can be written as

\begin{equation}
{\Delta B_r(t,\theta)\over\Delta t}=b_r(\theta)\omega_c\cos\omega_ct+
{b_r(\theta)\over A}\omega_\delta\cos(\omega_\delta t+\varphi).
\label{deriv}
\end{equation}

When $\omega_\delta \cong 10 \omega_c$, $A  \cong 2$ as it took
place in cycle 20, then the low-frequency term dominates in the
expression for $B_r$ (equa\-tion~\ref{field}) and the
high-frequency term prevails in the expression for $\Delta B_r/
\Delta t$ (equation~\ref{deriv}). Thus, presuming the existence of
the double magnetic cycle, one finds that the low-frequency term
prevails in the $B_r$-component, while the high - frequency term
dominates in $\Delta B_r/ \Delta t$.

Contour plots of $\Delta B_r/\Delta t$ as a function $\mu $(or
$\cos \theta$) and time, {\it t}, measured in Carrington rotation
are represented in Figure~1d for $\Delta t=6P$. During the cycles
21 and 22 the zones of increasing and decreasing strength of the
surface magnetic field appear in both the $N$ and $S$ hemispheres
(see Figure 1d). The width of these zones is approximately 2
years. This result coincides with investigations done earlier
based on Wilcox and Mount Wilson synoptic maps. Corresponding
power spectra for $\Delta B_r/\Delta t$  show that the component
with period $T=1.5 - 2.5$ years is dominant in both hemispheres
(\cite {Ben95}, \cite {Ben96}). These results are the basis of our
suggestion of a two component of solar cycle.

\section{\bf Model}

The possible explanation of the double magnetic cycle is that the
magnetic fields are generated by Parker's dynamo acting in
convective zone. The low-frequency component is generated at the
base of the convective zone due to large scale radial shear
$\partial \Omega /\partial r$; $\Omega$ is the angular velocity.
The high-frequency component may be generated in subsurface
regions due to latitudinal shear $\partial \Omega /
\partial \theta$ or due to radial shear. The
recent investigations of solar interior rotation  show a
significant radial gradient of angular velocity exists in
subsurface of the convective zone together with the latitudinal
gradient of the angular velocity (\cite {Schou98}). For simplicity
we use only latitudinal shear for generation of the high-frequency
component. Cartesian coordinates are employed, with $x$ denoting
the radial, $y$ the azimuthal and $z$ the latitudinal coordinate.
We consider axisymmetrical solutions ($\pd{ }\varphi=0$).

At the base of the convection zone  turbulence is suppressed by strong
magnetic field (\cite{Par93}) and, therefore,  diffusivity in
the first layer $\eta_1$
will could be less then $\eta_2$ in the second layer.
\medskip
The axisymmetrical mean magnetic field is decomposed into toroidal
and poloidal  parts. The following  set of equations describe
evolution of the magnetic field in thin layers at the levels $x_1$
and $x_2$:

\begin{eqnarray}
   \left\{
   \begin{array}{lll}
   \pd {A_{y1}}t &=&(\alpha_1+C_1)B_{y1} + \eta_1 \Delta A_{y1} \\
         & & \\
   \pd {B_{y1}}t &=& -\pd {A_{y1}}z G_x +\eta_1 \Delta B_{y1}\\
                 & & \\
   \pd{C_1}t      &=& -\nu C_1 + p A_{y1} B_{y1}
   \end{array}
   \right.
\label{x1}
\end{eqnarray}
\begin{eqnarray}
   \left\{
   \begin{array}{lll}
   \pd {A_{y2}}t &=&(\alpha_2+C_2)B_{y2} + \eta_2 \Delta A_{y2} \\
         & & \\
   \pd {B_{y2}}t &=& \pd {A_{y2}}x G_z +\eta_2 \Delta B_{y2} \\
                 & & \\
   \pd{C_2}t      &=& -\nu C_2 + p A_{y2} B_{y2}
   \end{array}
   \right.
\label{x2}
\end{eqnarray}

\noindent where $B_y$ is the toroidal magnetic field and  $A_y$ is
the azimuthal vector potential which gives poloidal field, $\eta$
is diffusivity and $\alpha$ is a kinetic helicity, $C$ is the
variable part of kinetic helicity. The first and the second
equations in the both systems (\ref{x1}) and (\ref{x2}) are a
generation of Parker's equations for large-scale radial shear
$G_x=\pd{V_y}x$ and large-scale latitudinal shear $G_z=\pd{V_y}z$
correspondingly. We have used the third equation in  the both
systems $\pd{C}t=-\nu C + p A B$ according to Kleeorin and
Ruzmaikin (1982) in a simple form, for feedback of the magnetic
field on the helicity. In these equations $\nu
> 0$ and $p <0 $ are parameters of the feedback. These two
non-linear differential equation systems describe the evolution of
two independent sources of the magnetic field. In the frame of
this model it is difficult to explain observed variations of the
high-frequency component. Therefore, we assume that the erupted
low-frequency magnetic field can influence the physical conditions
in the region where high-frequency component operates, through
modifying the helicity there. This allows the high-frequency
component to vary with time. In this case, the equation for $C_2$
becomes $\pd{C_2}t = -\nu C_2 + p( A_{y2} + a A_{y1}) (B_{y2} + a
B_{y1})$, where the parameter $a$ captures the feedback of the
low-frequency component of the magnetic field on the helicity in
regions where the high-frequency component is found. Following
Weiss et al. (1984) the solutions of these equations were found as
$A_1=a(t) e^{ikx}, B_1=b(t) e^{ikx}$ and $C_1=c_o(t) + c(t)
e^{2ikx}$, Im $c_o = 0$ for $x_1$ level and the same expressions
for $x_2$ level where $x$ is replaced by $z$. $A, B, C$ are
complex functions as $A=a_1+ia_2$, $B=b_1+ib_2$.

It is convenient to present the system in dimensionless units.
Since there will be an underlying periodicity in all calculations,
we use those periods as relative units of time. In relative units
$t \longrightarrow t^{'}=\eta_2 k^2 t\alpha_o$,
$\alpha\longrightarrow \alpha^{'}=\alpha/\eta_2 k^2 \alpha_o$,
$\sigma =\eta1/\eta2$,
  $C\longrightarrow C^{'}=C/\eta_2 k^2 \alpha_o,
 G_z\longrightarrow G_z^{'}=G_z/\eta_2 k \alpha_o$  and
$G_x\longrightarrow G_x^{'}=G_x/\eta_2 k \alpha_o$, where
$\alpha_o$ is some mean value of the kinetic helicity. Therefore,
$G_x$ and $G_z$ are represented as dynamo numbers in our model.
Therefore, set of the partial differential equations reduced to
the following equations were then solved numerically.

\begin{eqnarray}
   \left\{
   \begin{array}{rcl}
   \dot{A}_1     &=& (\alpha_1+C_1) B_1 -  \sigma A_1 +\frac{1}2 B_1^{*} \; C_1 \\
   \dot{B}_1     &=& -iA_1^{*} \; G_x - \sigma B_1 \\
   \dot{C}_1       &=& -\nu C_1 + p A_1 B_1
   \end{array}
   \right.
\label{x1a}
\end{eqnarray}
\begin{eqnarray}
   \left\{
   \begin{array}{rcl}
   \dot{A}_2      &=& (\alpha_2+C_2) B_2 -  A_2 +\frac{1}2 B_2^{*} \; C_2 \\
   \dot{B}_2      &=& i A_2 G_z  -  B_2 \\
   \dot{C}_2      &=& - \nu C_2 + p (A_2 + a A_1) (B_2 + a B_1)
   \end{array}
   \right.
\label{x2a}
\end{eqnarray}

\placefigure{fig2}
\begin{figure}
\epsscale{1.2} \hspace*{-2cm}\plotone{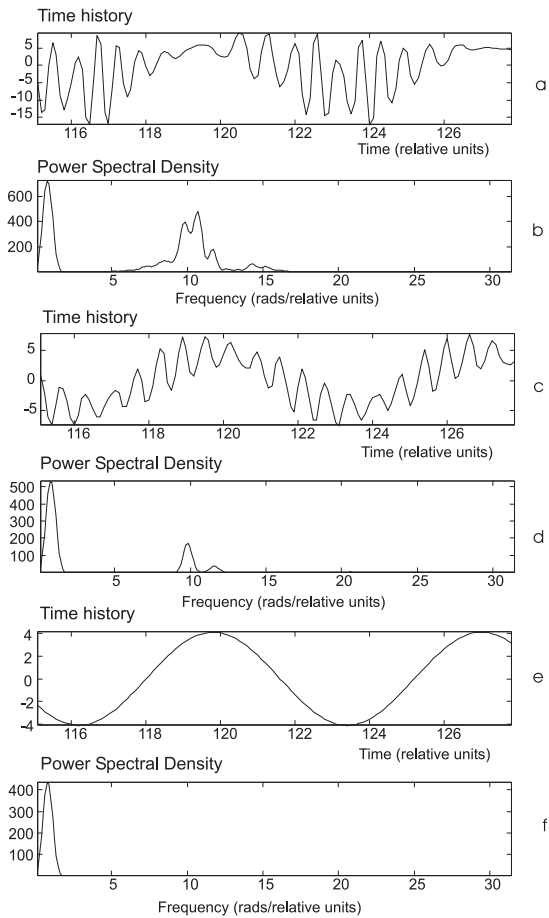} \caption{ (a)~Real
part of $A=A_1+A_2$ as function time for independent systems
($a=0$). (b)~Power Spectrum for case (a) in relative units.
(c)~Real part of $A=A_1+A_2$  as function time for $a=0.17$.
(d)~Power Spectrum for case (c) in relative units. (e)~Real part
of $A=A_1+A_2$  as function time for $a=0.2$. (f)~Power Spectrum
for case (e) in relative units.
 }
\end{figure}


We have investigated solutions of these two non-linear systems as
functions of the parameter $a$. Results for three principal cases,
$a = 0$, $0< a < 2$ and $a = 2$ are represented in Figure~3, where
$\alpha_1 = \alpha_2 = 1$, $\nu = 0.5$, $p = -1$, $\sigma = 0.1$,
$G_x = 0.3$, $G_z = 20$, which are in the reasonable range for the
Sun. If $a=0$ we have two independent sources of the magnetic
field which together generate low-frequency and high-frequency
signatures (Fig.~2~a,~b). For a weak interaction between these
sources ($0 < a < 0.2$) we still get a low and high frequencies
response. (Fig.~2~c,~d). Moreover, the high- frequency component
becomes more regular, and consequently is decreased when $a$
increases. With further increase of the feedback  of the
low-frequency magnetic field on the helicity near the top surface
the high-frequency component disappears and  only a low-frequency
mode regime is established (Fig.~2~e,~f). Therefore, in the case
of weak interaction between two sources of magnetic field it is
possible to obtain a stable double magnetic cycle.

\section{\bf Conclusions}

 Our model simulates the double magnetic cycle and temporal
variations of the biennial cycles from one 11-year cycle to
another. According to the magnetograph data (\cite {Ben96}) the
2-year component more clearly appeared in the northern hemisphere
in cycle 20 and in the southern hemisphere in cycle 21. In cycle
22 it was present in both hemispheres but was of lower amplitude.

In our model, a reduced high-frequency component occurs  when the
erupted magnetic field of the main (Hale's) cycle imposes the
helicity in the regions of generation of the high-frequency
component. The high-frequency component is more pronounced when
the effect of erupting magnetic fields of the main cycle on
helicity in this region is small. Thus, this simple model provides
a qualitative explanation of the double magnetic cycle. The next
step to a quantitative model is in development of a 2D model in
spherical geometry, since this simple model gives only a
qualitative picture of the two-components of the solar magnetic
cycle.

\noindent
{\bf Acknowledgments.}

\noindent I am grateful to Dr. J.W.Harvey for providing us the
Kitt Peak magnetograph data and to Drs. P.A.Gilman, J.T.Hoeksema,
A.G.Kosovichev and P.H.Scherrer for useful discussions and Russian
Federal Programme ``Astronomy'', grant 1.5.3.4.

\end{document}